\documentclass[aps,prl,twocolumn,preprintnumbers,10pt,showpacs]{revtex4-1}

\usepackage{amsmath,bm,amssymb}

\usepackage{graphicx}
\usepackage{psfrag}
\usepackage[caption=false]{subfig}
\usepackage{color,xcolor}
\definecolor{urlblue}{rgb}{0.2,0.4,0.7}
\definecolor{citegreen}{rgb}{0,0.6,0.2}
\definecolor{linkred}{rgb}{0.9,0.2,0.1}
\usepackage{hyperref}
\hypersetup{
colorlinks=true, citecolor=citegreen, linkcolor=blue,urlcolor=urlblue}

\usepackage{slashed}
\usepackage{array}
\newcolumntype{P}[1]{>{\centering\arraybackslash}p{#1}}

\allowdisplaybreaks

\def\be{\begin{equation}}
\def\ee{\end{equation}}
\def\nn{\nonumber\\}

\def\Ao{A_1^I}
\def\At{A_2^I}
\def\Ath{A_3^I}
\def\fo{f_1^I}
\def\ft{f_2^I}
\def\fth{f_3^I}

\def\Bo{B_1^I}
\def\Bt{B_2^I}
\def\Bth{B_3^I}
\def\goot{\bar{\mathcal{G}}_{1,\rm{SJ}}^{I,1}}
\def\gott{\bar{\mathcal{G}}_{1,\rm{SJ}}^{I,2}}
\def\gtot{\bar{\mathcal{G}}_{2,\rm{SJ}}^{I,1}}
\def\gthot{\bar{\mathcal{G}}_{3,\rm{SJ}}^{I,1}}
\def\gttt{\bar{\mathcal{G}}_{2,\rm{SJ}}^{I,2}}
\def\gotht{\bar{\mathcal{G}}_{1,\rm{SJ}}^{I,3}}
\begin{document}

\preprint{IMSc/2018/05/04}

\title{Gluon jet function at three loops in QCD
}

\author{Pulak Banerjee$^{a,b}$}\email{bpulak@imsc.res.in}
\author{Prasanna K. Dhani$^{a,b}$}\email{prasannakd@imsc.res.in}
\author{V. Ravindran$^{a,b}$}\email{ravindra@imsc.res.in}

\affiliation{$^a$ The Institute of Mathematical Sciences, Taramani,
 Chennai 600113, India \\ $^{b}$ Homi Bhabha National Institute,
 Training School Complex, Anushakti Nagar, Mumbai 400085, India}


\begin{abstract}
We present here the first result on the three-loop gluon jet function in perturbative QCD.  Using 
the three-loop coefficient functions \cite{Vermaseren:2005qc,Soar:2009yh} 
for deep-inelastic scattering via the exchange of a virtual photon that
couples to quarks or a scalar that couples to gluons and employing the    
KG equation, renormalization group invariance and factorization theorem, we obtain both the quark and the gluon 
jet functions up to the three-loop level. The former agrees with the recent result \cite{Bruser:2018rad}.  
These jet functions being universal ingredients  
in the SCET framework, will play an important role in the phenomenological studies 
at the Large Hadron Collider, 
such as resummation of jet observables and also in N-jettiness subtraction method.
\\

\end{abstract}


\maketitle
\section{I.\,INTRODUCTION}
The Standard Model (SM) is currently being tested to unprecedented accuracy at the Large Hadron Collider (LHC).
In order to  achieve such feats, it is important to accurately measure various observables 
and compare them with the precise theoretical predictions that uses state-of-the-art modern 
techniques.  In addition such studies are of paramount importance to understand
the SM background to constrain any physics beyond the SM.  In particular, processes such as
production of lepton pairs, vector bosons and multi-jets at the LHC play an important
role in these studies. In addition, understanding of the jets will shed light on
the underlying structure of QCD dynamics.
The hadronic cross sections in high-momentum transfer processes such as the 
lepton-nucleon deep-inelastic scattering (DIS) or in the Drell-Yan processes factorize into
 hard and soft virtual (SV) parts. The hard part is associated with the physics of large momentum transfer $Q$; the soft part of the SV cross section describes the dynamics associated with emission of soft and collinear partons. Soft-collinear effective theory (SCET) \cite{ Bauer:2000ew,Bauer:2000yr,Bauer:2001ct,Bauer:2001yt,Bauer:2002nz,Beneke:2002ph} captures the physics of soft
and collinear dynamics of these processes at high energies through the soft and jet functions. 
The jet functions explains the 
propagation of collinear partons  inside jets and the soft functions mediate low-energy interaction
between jets.  They are  important components of various observables that can be measured 
at the colliders.   In SCET, the quark and gluon soft as well as jet functions have been computed to higher orders in perturbation theory.  The quark and the gluon soft functions are
already known to the three-loop level; see \cite{Li:2014bfa} and references therein.  For the quark jet function, results were known
up to the two-loop level for some time (see \cite{Bauer:2003pi,Bosch:2004th,Becher:2006qw}), and 
the results at three loops were computed recently in \cite{Bruser:2018rad}.
On the other hand, only one-~\cite{Becher:2009th} and two-loop~\cite{Becher:2010pd} results are known for the gluon jet function.  Precise measurements of various observables at high-energy experiments demand 
accurate theoretical predictions from the SM.  The latter obtained using sophisticated 
methods in turn shed light on the theoretical structure of the underlying dynamics.  In the past,
such computations have led to better understanding of the perturbative structure of various
observables in terms of process-independent functions such as soft and jet functions 
and their anomalous dimensions.  Owing to the fact that these quantities are universal, they form
the building blocks for many interesting observables that probe physics at short distances. 
Our goal in this article is to demonstrate an intriguing connection between jet functions in SCET and coefficient functions of DIS cross sections computed in perturbative QCD. We exploit this novel connection to obtain quark and gluon jet functions up to three-loop level using the known state-of-the-art three-loop coefficient functions in the context of DIS.
In particular, we use the coefficient functions computed in \cite{Soar:2009yh} 
which uses a scalar that couples to gluons 
to probe the short distance structure of the hadron and  obtain the three-loop gluon 
jet function. To achieve this, we exploit 
the KG equation, renormalization group invariance, factorization theorem and use various 
three-loop results.  We use the framework developed in  
\cite{Ravindran:2005vv,Ravindran:2006cg,Ravindran:2006bu} which describes resummation of soft gluons to
all orders in QCD perturbation theory. For the recent results on the soft gluon resummation of the rapidity distribution of Higgs and lepton pair production in Drell-Yan, see~\cite{Banerjee:2017cfc,Banerjee:2018vvb}.   In \cite{Ravindran:2006cg}, it was shown that 
DIS cross section in the threshold limit
factorizes into the square of ultraviolet (UV) renormalized form factor, soft plus jet function and 
the mass factorization kernels.   The soft plus jet function has an universal structure and it 
depends only on the nature of external states namely quark or gluon states.   In addition,
it satisfies KG type differential equation similar to the form factors.  
Factorization properties of the cross section and renormalization group invariance can be
used to unravel the structure of these soft plus jet function to all orders in perturbation theory.
Upon identifying the finite part of the square of the form factor with the matching hard coefficient
in the SCET framework, the finite part of soft plus jet function can be shown to 
coincide with the corresponding jet function.  The latter identification is simply due to 
the process-independent nature of both the soft plus jet function and the jet function.  
We first demonstrate that the quark jet function known up to the three-loop level agrees with
the finite part of quark soft plus jet function obtained from the coefficient function \cite{Vermaseren:2005qc}
and proceed to obtain the corresponding gluon jet function from \cite{Soar:2009yh}. 


\section{II.\,THEORETICAL FRAMEWORK AND RESULTS}
The inclusive cross section for the scattering of a lepton with a hadron in the DIS is given by
\begin{eqnarray}
\label{eq1}
\sigma^{I}(x, Q^2) &=& \sigma^I_B (\mu_R^2) \sum_{a=q,\overline q,g } \int_{x}^{1} \, \frac{dz}{z} f_{a}\left(\frac{x}{z},\mu_{F}^2\right)  \nonumber \\
&&\times \Delta^{I}_{a}(a_s,z, Q^2, \mu_{R}^2, \mu_{F}^2).
\end{eqnarray}
The scaling variable at the hadronic level is given by $x=\frac{-q^2}{2 P.q}$. $P$ and $q$ are the momenta of the hadron and the 
intermediate off-shell particle.  For DIS we have $Q^2=-q^2$. 
The parton distribution function is given by $f_{a}\left(y,\mu_{F}^2\right)$ with the momentum fraction $y$ 
of the hadron at the factorization scale $\mu_F$. 
The particle probing the structure of the hadron can be a photon ($I=q$) or the scalar particle
($I=g$). The scaling variable at the partonic  level is $z = \frac{-q^2}{2p.q}$ where $p$
is the fraction of the parent hadron momentum. The perturbatively computed UV and infrared (IR)
finite part of the partonic cross section, called the coefficient function, 
is given by $\Delta^{I}_{a}(a_s, z, Q^{2}, \mu_{R}^2, \mu_{F}^2)$, where  $\mu_R$  denotes the UV renormalization scale. 
The Born contribution $\sigma^I_B$ is chosen in such a way that $\Delta^I_a$ to lowest order
in perturbation theory is equal to $\delta(1-z)$.
In dimensional regularization, 
the UV renormalized strong coupling constant is $a_s =  g_s^2(\mu_R^2)/16 \pi^2 $
is written in 
terms of bare coupling constant  $\hat a_s = \hat g_s^2/16 \pi^2$  as $
\hat a_s(\mu^2) S_{\epsilon}= a_s (\mu_R^2)(\mu^2/\mu_R^2) ^{\epsilon/2}  Z(a_s(\mu_R^2))$. 
%
The scale $\mu$ keeps  $\hat g_s$ dimensionless in $n = 4 + \epsilon$ space-time dimensions.
$S_\epsilon$ is the spherical function defined as $S_\epsilon = \exp \big[ (\gamma_E-\ln 4 \pi) \frac{\epsilon}{2} \big]$,  where $\gamma_E$ is the Euler-Mascheroni constant. $Z(a_s(\mu_R^2))$ is the coupling constant renormalization term.

The infrared safe coefficient functions $\Delta_a^I$ get contributions from both soft gluons as well as 
from hard partons. We write them as sum of contributions from soft plus virtual and the remaining hard part:
\begin{align}
\label{eq3}
\Delta^I_{a} (a_s,z,Q^2,\mu_F^2,\mu_R^2)  = & 
\,\, \Delta^{I,{\rm hard}}_{a}(a_s, z,Q^2,\mu_F^2,\mu_R^2) \nn
 & +  \Delta^{I,\text{SV}}(a_s, z,Q^2,\mu_F^2,\mu_R^2).
\end{align}
The SV part of the cross section consists only of plus 
distributions ${\cal D}_i(z)=\Big[\frac{\ln^i(1-z)}{1-z}\Big]_+$ and $\delta(1-z)$. In the soft limit 
$z \rightarrow 1$ these distributions 
give dominant contributions to the hadronic cross section after they are convolved with the parton distribution functions as expressed through~\eqref{eq1}. 
As discussed in \cite{Ravindran:2005vv}
the $\text{SV}$ part of the coefficient function  
can be shown to factorize in terms of  
the square of the UV renormalized virtual contributions, soft plus jet distribution function
appropriately convoluted with mass factorization kernels as
\begin{align}
\label{eq4}
\Delta^{I,\rm SV}( z,Q^2) = &  (Z^{I} (\hat a_s,\mu_R^2, \mu^2, \epsilon))^2 \big| \hat {\cal F}^{I}(\hat a_s, Q^2,\mu^2,\epsilon)\big|^2
  \nn & \delta(1-z) \otimes {\cal C} e^{2 \Phi^{I}_{\rm {SJ}}(\hat a_s, Q^2, \mu^2, z,\epsilon)}  \nn
& \otimes \Gamma_{II} ^ {-1}(\hat a_s, \mu^2, \mu^2_{F}, z, \epsilon)\,.  
\end{align}
The symbol ${\cal C}$ denotes convolution and its operation on the exponential of a function $f(z)$ can be found in \cite{Ravindran:2005vv}. 
%
%
and  
 $\otimes$ indicates the Mellin convolution,  which convolutes  with respect to the variables $z$. As we are interested in evaluating the $\text{SV}$ part of the cross sections, we neglect all the regular 
 functions that come from different convolutions. 
 In \eqref{eq4}, the overall renormalization constant, $Z^{I} (\hat a_s,\mu_R^2, \mu^2, \epsilon)$, for 
$I=g$ can 
be obtained from that of Higgs-gluon effective 
operator obtained in the large top quark mass limit and the exact form can be written 
in term of the anomalous dimension $\gamma_g$, which is known to the three-loop level \cite{Chetyrkin:1997un}. 
 For $I=q$, one finds   $Z^q (\hat a_s,\mu_R^2,\mu^2, \epsilon) =1$ to all orders. 
 $\Gamma_{II} (\hat a_s, \mu^2, \mu^2_{F}, z, \epsilon)$ is the mass factorization kernel,
 known fully up to the three-loop
level \cite{Moch:2004pa,Vogt:2004mw} and in the large $n_{f}$ limit at the four-loop level  \cite{Davies:2016jie}. 
The quantity $\hat {\cal F}^{I}(\hat a_s, Q^2,\mu^2,\epsilon)$ is the bare form factor which satisfies the 
KG equation, where the latter is a consequence of factorization, gauge and renormalization group invariances \cite{Sudakov:1954sw,Mueller:1979ih,Collins:1980ih,Sen:1981sd}. Its general solution up to 
four loops  can be found in \cite{Moch:2005id,Ravindran:2005vv}
in terms of the universal three-loop 
cusp ($A^{I}$)  \cite{Moch:2004pa,Vogt:2004mw,Catani:1989ne,Catani:1990rp,Vogt:2000ci}, collinear ($B^{I}$) 
 \cite{Moch:2004pa,Vogt:2004mw}, soft ($f^{I}$) 
 \cite{Ravindran:2004mb,Vogt:2004mw} 
anomalous dimensions known to the three-loop level and some form factor dependent constants.
The quantity $\Phi^{I}_{\rm {SJ}}(\hat a_s, Q^2, \mu^2, z,\epsilon)$ in \eqref{eq4} is the soft plus jet
 distribution function
which contains singular as well as finite parts due to soft gluons and collinear parton emissions.
In~\cite{Ravindran:2005vv,Ravindran:2006cg}, it was shown that by demanding finiteness of  $\Delta^{I,\rm SV}$,  $\Phi^{I}_{\rm SJ}(\hat{a}_s, Q^2, \mu^2, z, \epsilon)$ can also be shown to
satisfy a Sudakov-type differential equation  namely(we suppress the arguments of $\Phi^{I}_{\rm SJ}$ for brevity),
\begin{equation}
\label{eq6}
 Q^2 \frac{d}{dQ^2} \Phi^I_{\rm SJ}  = \frac{1}{2} \Big[ \bar{K}^I (\hat{a}_s, \frac{\mu_R^2}{\mu^2}, z,
\epsilon ) + \bar{G}^I_{\rm SJ} (\hat{a}_s, \frac{Q^2}{\mu_R^2}, \frac{\mu_R^2}{\mu^2}, z, \epsilon ) \Big] \, .
\end{equation}
The terms $\bar{K}^I$ and $\bar{G}^I$ are such that the former contains
poles as $\epsilon \rightarrow 0$ while the latter is finite in the same limit.
Their expressions have similar forms as those of the 
corresponding ones that appear in form factors.
The solution to the above equation is found to be
\begin{eqnarray}
\label{eq7}
\Phi^I_{\rm SJ} = \sum_{i=1}^\infty {\hat a}_s^i S_\epsilon^i \left({Q^2 (1-z) \over \mu^2}\right)^{i {\epsilon \over 2}}
 {i \epsilon \over 2(1-z)} \hat \phi^{I,(i)}_{\rm SJ} (\epsilon)
\end{eqnarray}
where $\hat \phi^{I,(i)}_{\rm SJ}(\epsilon) = \big[\bar{K}^{I,(i)}(\epsilon) 
+ \bar{G}^{I,(i)}_{\rm SJ}(\epsilon)\big]/i \epsilon$.
Expressing $\bar{K}^I = \sum_{i=1}^\infty \hat a_s^i \left({\mu_R^2 \over \mu^2}\right)^{i {\epsilon \over 2} }
S_\epsilon^i \bar{K}^{I,(i)}$,  
the coefficients $\bar{K}^{I,(i)} (\epsilon)$ 
can be expressed in terms of $A^I_i$, beta function of QCD $\beta_i$  \cite{Tarasov:1980au} 
and $\bar{G}^{I,(i)}_{\rm SJ}(\epsilon)$ given by
\begin{eqnarray}
\label{eq8}
\sum_{i=1}^\infty \hat a_s^i \left( {Q_z^2  \over \mu^2}\right)^{i {\epsilon \over 2}}
S_\epsilon^i \bar{G}^{I,(i)}_{\rm SJ}(\epsilon) =
\sum_{i=1}^\infty a_s^i(Q_z^2 ) 
\bar{{\cal G}}^I_{i,\rm SJ}(\epsilon)
\end{eqnarray}
with $Q_z^2=Q^2 (1-z)$, can be
expressed in terms of the $B^{I}-, f^{I}-$ and $\epsilon$-dependent part from lower order
in the following way:
$\bar{{\cal G}}^I_{i,\rm SJ} =  -(B_{i}^{I} + f_{i}^{I})  + C^{I}_i + \sum_{k=1}^{\infty} \epsilon^k  
\bar{{\cal G}}^{I,k}_{i,\rm SJ},
$
where the constants $C_{i}^{I}$ up to three-loop order are given by
$C_{1}^{I} = 0\, , 
C_{2}^{I} = - 2 \beta_{0} \bar{{\cal G}}^{I,1}_{2,\rm SJ} \, ,
C_{3}^{I} = - 2 \beta_{1} \bar{{\cal G}}^{I,1}_{1,\rm SJ} - 2 \beta_{0} \left(\bar{{\cal G}}^{I,1}_{2,\rm SJ} + 2 \beta_{0} \bar{{\cal G}}^{I,2}_{1,\rm SJ}\right)\, .
$
The $z$-independent constants $\bar{{\cal G}}^{I,k}_{i,\rm SJ}$ are determined 
from the explicit computation of the SV coefficient functions $\Delta^{I,\rm SV}(z,Q^2)$.

Computation of the coefficient functions $\Delta^I_a$  
in perturbative QCD plays an important role in understanding the structure of hadrons.
In the DIS process, the cross section factorizes into hadronic 
and the leptonic parts and the former can be computed 
by using operator product expansion in the Bj\"orken limit.
Using various symmetries, the hadronic part  
can be expressed in terms of the structure functions $F_1(x,Q^2)$ and $F_2(x,Q^2)$. 
These functions factorize into calculable coefficient functions $c_i(x,Q^2,\mu_F^2)$, $i=1,2$ 
and nonperturbative parton distribution 
functions $f_a(x,\mu_F^2)$, $a=q, \bar{q}, g$.
Applying the optical theorem, 
one relates the DIS cross section to the imaginary part of the forward scattering amplitude, 
where a virtual photon scatters off a
nucleon. 
This forward scattering amplitude can be written in terms of coefficient functions $c_i$, 
where the latter can be computed by expanding in a perturbative series of the strong coupling constant. 
Computation of the higher-order coefficient functions 
\cite{vanNeerven:1991nn,Zijlstra:1992kj,Zijlstra:1991qc,Zijlstra:1992qd,Hamberg:1990np}
along with the higher order splitting functions \cite{Moch:2004pa,Vogt:2004mw} 
and the precise measurements at DIS experiments were used to extract $F_1$ and $F_2$ accurately. 
Using the nonsinglet part of the quark coefficient function $c_{2,q}$ computed up to three loops 
\cite{Vermaseren:2005qc} and $c_{\phi,g}^3$ computed using the off-shell scalar DIS process in 
\cite{Soar:2009yh}, we can extract $\Delta^{q,\rm SV}(z,Q^2)$ and $\Delta^{g,{\rm SV} }$, respectively,  up to three-loops.   Using these results and the known three-loop results for $A,B,f$ and the form factor dependent
constants, we can determine $\bar{{\cal G}}^{I,k}_{i,{\rm SJ}}$ to desired accuracy in $\epsilon$.

At the hadron colliders, the jets of quarks and gluons~\cite{Sterman:1977wj, Salam:2007xv, Cacciari:2008gp} 
capture the properties of QCD and provide insight into the IR structure of QCD processes.  
Jets are useful to study both SM as well as BSM processes. 
The definition of jets and its various properties play an important role in various new physics
search scenarios.  SCET provides a suitable framework to study the scattering or decay processes
involving jet final states with invariant masses having large hierarchy with the center-of-mass energy
of the process.  The corresponding observables can be factorized in terms of certain process
-dependent functions such as hard functions and process-independent soft and jet functions.       
Jet functions can result from quark or gluon radiating jet of collinear partons. 
The jet functions are important ingredients of SCET factorization for many processes with 
quark and gluon initiated states.
The massless quark jet function in SCET at one loop was computed in \cite{Bauer:2003pi,Bosch:2004th}; for gluon see \cite{Becher:2009th}.
The tw- loop results for the quark can be found in \cite{Becher:2006qw}; for the gluon, see \cite{Becher:2010pd}. 
The three-loop quark jet function have been recently computed in  \cite{Bruser:2018rad}.

If we apply SCET formalism to the DIS process for the cases with $I=q,g$, we can identify the 
UV and IR finite parts of the form factor and soft plus jet function with the process-dependent matching coefficient
and the jet function of SCET, respectively.
Note that only one massless parton initiates the hard process in each case, (for $I=q$, 
quark/antiquark scatters of the virtual photon in the hard process and for $I=g$, gluon scatters of
off-shell scalar), there will be only one jet function in each case and hence it is straightforward
to identify it with the finite part of soft plus jet function.

The soft plus jet function $\Phi^I_{\rm SJ}$ can be factorized into  
part containing IR poles in $\epsilon$ and a part containing the finite terms in the limit 
$\epsilon \rightarrow 0$, that is  
\be
\label{eq11}
{\cal C} e^{2 \Phi^{I}_{\rm SJ}} = {\cal Z}^{I} \, \otimes {\cal C} e^{2 \Phi^{I, \rm{fin}}_{\rm SJ}}
\ee
where ${\cal Z}^{I}$ contains only IR poles in $\epsilon$ and can be expanded as
\be
\label{eq12}
 {\cal Z}^{I} = \delta(1-z) + \sum_{i=1}^{n} \sum_{j=1}^{2\, i} a_s^{i}\, \frac{{\cal Z}^{I}_{i\,j}}{\epsilon^{j}}.
\ee 
The coefficients ${\cal Z}^{I}_{i\,j}$ contain $A^{I}$, $B^{I}, f^{I}$, $\delta(1-z)$ and ${\cal D}_i(z)$ and it
reads as 
\begin{align}
\label{eq12}
\mathcal{Z}^{I}_{11}\big|_{\mathcal{D}_{0}} &=2\Ao,\,
\mathcal{Z}^{I}_{11}\big|_{\delta} = -2\left(\Bo+\fo\right),\,
\mathcal{Z}^{I}_{12}\big|_{\delta} =4\Ao,\,
\nn
\mathcal{Z}^{I}_{21}\big|_{\mathcal{D}_{0}} &=\At,\,
\mathcal{Z}^{I}_{21}\big|_{\delta} =-\left(\Bt+\ft\right),\,
\mathcal{Z}^{I}_{22}\big|_{\mathcal{D}_{1}} =4{\Ao}^2,\,
\nn
\mathcal{Z}^{I}_{22}\big|_{\mathcal{D}_{0}} &=-4\Ao\left(\Bo+\fo\right)+2\beta_0\Ao,\,
\mathcal{Z}^{I}_{22}\big|_{\delta} =
\nn
&2\left( \Bo+\fo\right)^2
+\At-2\beta_0\left(\Bo+\fo \right)
-\frac{1}{3}\pi^2{\Ao}^2,\,
\nn
\mathcal{Z}^{I}_{23}\big|_{\mathcal{D}_{0}} &=8{\Ao}^2,\,
\mathcal{Z}^{I}_{23}\big|_{\delta} =-8\Ao\left( \Bo+\fo\right)+6\beta_0\Ao,\,
\nn
\mathcal{Z}^{I}_{24}\big|_{\delta} &=8{\Ao}^2,\,
\mathcal{Z}^{I}_{31}\big|_{\mathcal{D}_{0}} =\frac{2}{3}\Ath,\,
\mathcal{Z}^{I}_{31}\big|_{\delta} =-\frac{2}{3}\left(\Bth+\fth \right),\,
\nn
\mathcal{Z}^{I}_{32}\big|_{\mathcal{D}_{1}} &=4\Ao\At,\,
\mathcal{Z}^{I}_{32}\big|_{\mathcal{D}_{0}} =-2\At\left(\Bo+\fo \right)
\nn
&-2\Ao\left(\Bt+\ft \right)
+\frac{4}{3}\beta_1\Ao+\frac{4}{3}\beta_0\At,\,
\nn
\mathcal{Z}^{I}_{32}\big|_{\delta} &=2\left(\Bo+\fo\right)\left( \Bt+\ft\right)+\frac{4}{9}\Ath
-\frac{4}{3}\beta_1\left( \Bo+\fo\right)
\nn
&-\frac{4}{3}\beta_0\left( \Bt+\ft\right)-\frac{1}{3}\pi^2\Ao\At,\,
\mathcal{Z}^{I}_{33}\big|_{\mathcal{D}_{2}} =4{\Ao}^3,\,
\nn
\mathcal{Z}^{I}_{33}\big|_{\mathcal{D}_{1}} &=-8{\Ao}^2\left( \Bo+\fo\right)+8\beta_0{\Ao}^2,\,
\nn
\mathcal{Z}^{I}_{33}\big|_{\mathcal{D}_{0}} &=4\Ao\left(\Bo+\fo \right)^2+6\Ao\At-8\beta_0\Ao\left(\Bo+\fo \right)
\nn
&+\frac{8}{3}\beta_0^2\Ao-\frac{2}{3}\pi^2{\Ao}^3,\,
\mathcal{Z}^{I}_{33}\big|_{\delta} =-\frac{4}{3}\left(\Bo+\fo \right)^3
\nn
&-2\At\left(\Bo+\fo \right)-4\Ao\left(\Bt+\ft \right)+\frac{32}{9}\beta_1\Ao
\nn
&+4\beta_0\left(\Bo+\fo \right)^2+\frac{20}{9}\beta_0\At-\frac{8}{3}\beta_0^2\left(\Bo+\fo \right)
\nn
&+\frac{8}{3}\zeta_3{\Ao}^3+\frac{2}{3}\pi^2{\Ao}^2\left( \Bo+\fo\right)-\frac{2}{3}\pi^2\beta_0{\Ao}^2,\,
\nn
\mathcal{Z}^{I}_{34}\big|_{\mathcal{D}_{1}} &=16{\Ao}^3,\,
\mathcal{Z}^{I}_{34}\big|_{\mathcal{D}_{0}} =-16{\Ao}^2\left(\Bo+\fo \right)+20\beta_0{\Ao}^2,\,
\nn
\mathcal{Z}^{I}_{34}\big|_{\delta} &=8\Ao\left(\Bo+\fo \right)^2+4\Ao\At-20\beta_0\Ao\left( \Bo+\fo\right)
\nn
&+\frac{88}{9}\beta_0^2\Ao-\frac{4}{3}\pi^2{\Ao}^3,\,
\mathcal{Z}^{I}_{35}\big|_{\mathcal{D}_{0}} =16{\Ao}^3,\,
\nn
\mathcal{Z}^{I}_{35}\big|_{\delta} &=-16{\Ao}^2\left(\Bo+\fo \right)+24\beta_0{\Ao}^2,\,
\mathcal{Z}^{I}_{36}\big|_{\delta} =\frac{32}{3}{\Ao}^3.
\end{align}
We now present the general expression for the jet function ${\cal C} e^{2 \Phi^{I, \rm{fin}}_{\rm SJ}} =
\delta(1-z) + \sum_{i=1}^\infty a_s^i J^{I}_{i} \big |_{k}$ up to three loops where $J^{I}_{i} \big |_{k}$ represent the coefficients of $ {\cal D}_{j}(z), \delta$ for $j \le (2 i-1)$. We have set $Q^2 = \mu_R^2=\mu_F^2$ in our computation:
\begin{align}
J^{I}_1\big|_{\mathcal{D}_1}&=\Ao,\,
J^{I}_1\big|_{\mathcal{D}_0}=-\left( \Bo+\fo\right),\,
J^{I}_1\big|_{\delta}=2\goot,\,
\nn
J^{I}_2\big|_{\mathcal{D}_3}&=\frac{1}{2}{\Ao}^2,\,
J^{I}_2\big|_{\mathcal{D}_2}=-\frac{3}{2}\Ao\left( \Bo+\fo\right)-\frac{1}{2}\beta_0\Ao,\,
\nn
J^{I}_2\big|_{\mathcal{D}_1}&=\left( \Bo+\fo\right)^2+\At+2\Ao\goot+\beta_0\left( \Bo+\fo\right)
\nn
&-\frac{1}{6}\pi^2{\Ao}^2,\,
J^{I}_2\big|_{\mathcal{D}_0}=-\left( \Bt+\ft\right)
\nn
&-2\goot\left( \Bo+\fo\right)
-2\beta_0\goot+\zeta_3{\Ao}^2
\nn
&+\frac{1}{6}\pi^2\Ao\left( \Bo+\fo\right),\,
\nn
J^{I}_2\big|_{\delta}&=\gtot+2\left( \goot\right)^2+2\beta_0\gott-\zeta_3\Ao\left(\Bo+\fo \right)
\nn
&-\frac{1}{12}\pi^2\left(\Bo+\fo \right)^2-\frac{1}{720}\pi^4{\Ao}^2,\,
J^{I}_3\big|_{\mathcal{D}_5}=\frac{1}{8}{\Ao}^3,\,
\nn
J^{I}_3\big|_{\mathcal{D}_4}&=-\frac{5}{8}{\Ao}^2\left(\Bo+\fo \right)-\frac{5}{12}\beta_0{\Ao}^2,\,
\nn
J^{I}_3\big|_{\mathcal{D}_3}&=\Ao\left( \Bo+\fo\right)^2+\Ao\At+\goot{\Ao}^2
\nn
&+\frac{5}{3}\beta_0\Ao\left( \Bo+\fo\right)+\frac{1}{3}\beta_0^2\Ao-\frac{1}{6}\pi^2{\Ao}^3,\,
\nn
J^{I}_3\big|_{\mathcal{D}_2}&=-\frac{1}{2}\left(\Bo+\fo \right)^3-\frac{3}{2}\At\left( \Bo+\fo\right)
\nn
&-\frac{3}{2}\Ao\left(\Bt+\ft \right)-3\goot\Ao\left(\Bo+\fo \right)-\frac{1}{2}\beta_1\Ao
\nn
&-\frac{3}{2}\beta_0\left(\Bo+\fo \right)^2-\beta_0\At-4\beta_0\goot\Ao
\nn
&-\beta_0^2\left( \Bo+\fo\right)+\frac{5}{2}\zeta_3{\Ao}^3+\frac{1}{2}\pi^2{\Ao}^2\left(\Bo+\fo \right)
\nn
&+\frac{1}{4}\pi^2\beta_0{\Ao}^2,\,
\nn
J^{I}_3\big|_{\mathcal{D}_1}&=2\left( \Bo+\fo\right)\left(\Bt+\ft \right)+\Ath
+\gtot\Ao
\nn
&+2\goot\left( \Bo+\fo\right)^2+2\goot\left(\At+\goot\Ao\right)
\nn
&+\beta_1\left(\Bo+\fo \right)+2\beta_0\left( \Bt+\ft\right)+2\beta_0\gott\Ao
\nn
&+6\beta_0\goot\left(\Bo+\fo \right)+4\beta_0^2\goot-5\zeta_3{\Ao}^2\left( \Bo+\fo\right)
\nn
&-3\zeta_3\beta_0{\Ao}^2-\frac{5}{12}\pi^2\Ao\left(\Bo+\fo \right)^2-\frac{1}{3}\pi^2\Ao\At
\nn
&-\frac{1}{3}\pi^2\goot{\Ao}^2-\frac{1}{2}\pi^2\beta_0\Ao\left( \Bo+\fo\right)-\frac{1}{144}\pi^4{\Ao}^3,\,
\nn
J^{I}_3\big|_{\mathcal{D}_0}&=-\left( \Bth+\fth\right)-\gtot\left( \Bo+\fo\right)-2\goot\left(\Bt+\ft \right)
\nn
&-2\left(\goot\right)^2\left( \Bo+\fo\right)-2\beta_1\goot-2\beta_0\gtot
\nn
&-2\beta_0\gott\left(\Bo+\fo \right)-4\beta_0\left( \goot\right)^2-4\beta_0^2\gott
\nn
&+3\zeta_5{\Ao}^3+2\zeta_3\Ao\left( \Bo+\fo\right)^2+2\zeta_3\Ao\At
\nn
&+2\zeta_3\goot{\Ao}^2+3\zeta_3\beta_0\Ao\left( \Bo+\fo\right)
\nn
&+\frac{1}{12}\pi^2\left(\Bo+\fo\right)^3+\frac{1}{6}\pi^2\At\left( \Bo+\fo\right)
\nn
&+\frac{1}{6}\pi^2\Ao\left(\Bt+\ft \right)+\frac{1}{3}\pi^2\goot\Ao\left(\Bo+\fo \right)
\nn
&+\frac{1}{6}\pi^2\beta_0\left( \Bo+\fo\right)^2+\frac{1}{3}\pi^2\beta_0\goot\Ao-\frac{1}{3}\pi^2\zeta_3{\Ao}^3
\nn
&+\frac{1}{144}\pi^4{\Ao}^2\left( \Bo+\fo\right)+\frac{1}{72}\pi^4\beta_0{\Ao}^2,\,
\nn
J^{I}_3\big|_{\delta}&=\frac{2}{3}\gthot+2\goot\gtot+\frac{4}{3}\left( \goot\right)^3+\frac{4}{3}\beta_1\gott
\nn
&+\frac{4}{3}\beta_0\gttt
+4\beta_0\goot\gott+\frac{8}{3}\beta_0^2\gotht
\nn
&-3\zeta_5{\Ao}^2\left(\Bo+\fo \right)
-2\zeta_5\beta_0{\Ao}^2-\frac{1}{3}\zeta_3\left(\Bo+\fo \right)^3
\nn
&-\zeta_3\At\left(\Bo+\fo \right)
-\zeta_3\Ao\left( \Bt+\ft\right)
\nn
&-2\zeta_3\goot\Ao\left( \Bo+\fo\right)
-\zeta_3\beta_0\left(\Bo+\fo \right)^2
\nn
&-2\zeta_3\beta_0\goot\Ao+\frac{5}{6}\zeta_3^2{\Ao}^3
\nn
&-\frac{1}{6}\pi^2\left(\Bo+\fo \right)\left(\Bt+\ft \right)
-\frac{1}{6}\pi^2\goot\left(\Bo+\fo \right)^2
\nn
&-\frac{1}{3}\pi^2\beta_0\goot\left(\Bo+\fo \right)
+\frac{1}{3}\pi^2\zeta_3{\Ao}^2\left(\Bo+\fo \right)
\nn
&+\frac{1}{6}\pi^2\zeta_3\beta_0{\Ao}^2
-\frac{1}{360}\pi^4\Ao\left(\Bo+\fo \right)^2
\nn
&-\frac{1}{360}\pi^4\Ao\At-\frac{1}{360}\pi^4\goot{\Ao}^2
\nn
&-\frac{1}{72}\pi^4\beta_0\Ao\left(\Bo+\fo \right)
-\frac{29}{45360}\pi^6{\Ao}^3.
\end{align}
Note that the coefficients $J^{(m)}_{i,n}$ given in \cite{Bruser:2018rad} are nothing but our
$J^{I}_i|_k , k=\delta , {\cal D}_j(z)$ if we identify $m,i,n$ with $i,I,k$  
in $J^{(m)}_{i,n}$ and ${\cal L}_n({s/\mu_R^2})$ with ${\cal D}_j$; 
the anomalous dimension of the jet function can be written in terms of
collinear ($B^I$)  and soft ($f^I$) anomalous dimension, 
that is $\gamma_J^I = B^I + f^I$ where $I=q,g$.
The renormalization group equation satisfied by soft plus jet function is given by
\be
\label{eqRG}
\mu_R^2 \frac{d}{d \mu_R^2} J^{I} = \Gamma_{J}^{I} \otimes J^{I}
\ee
where $\Gamma_{J}^{I} = \left\{B^I + f^I - A^I \ln \left( \frac{Q^2}{\mu_R^2}\right)\right\} \delta(1-z) - A^I {\cal D}_0$.
We can get the logarithmic-dependent parts of $J^{I}$ through the above equation.
Rescaling $(1-z)$ by $Q^2/z\mu_R^2$, we can easily relate our $\Gamma^I_J$ with $\gamma^I_J(s,\mu_R^2)$
given \cite{Bruser:2018rad}

For $I=q$, our result ${\cal C} e^{2 \Phi^{q, \rm{fin}}_{\rm SJ}}$ matches with the three-loop quark 
jet function given in \cite{Bruser:2018rad} after identifying $s$ in the latter reference
without $z$ through $s=(p+q)^2 = Q^2 (1-z)/z$ and taking the soft limit $z\rightarrow 1$. 
Below, we present the gluon jet function up to three loops:
\begin{align}
J^g_1\big|_{\mathcal{D}_{1}} &= C_A\bigg[4 \bigg],\,\quad
J^g_1\big|_{\mathcal{D}_{0}} =C_A\bigg[ -\frac{11}{3}
\bigg]+n_f\bigg[ \frac{2}{3}\bigg],\,
\nn
J^g_1\big|_{\delta} &=C_A\bigg[\frac{67}{9}-\pi^2
\bigg]
+n_f\bigg[ -\frac{10}{9}
\bigg]\,,
\nn
J^g_2\big|_{\mathcal{D}_{3}} &=C_A^2\bigg[ 8\bigg],\, \quad
J^g_2\big|_{\mathcal{D}_{2}} =C_A^2\bigg[ -\frac{88}{3}
\bigg]+C_A n_f\bigg[ \frac{16}{3}\bigg],\,
\nn
J^g_2\big|_{\mathcal{D}_{1}} &=C_A^2\bigg[\frac{778}{9}-8\pi^2
\bigg]
+C_A n_f\bigg[ -\frac{56}{3}
\bigg]+n_f^2\bigg[ \frac{8}{9}\bigg],\,
\nn
J^g_2\big|_{\mathcal{D}_{0}} &=C_A^2\bigg[ -\frac{2570}{27}+32\zeta_3+11\pi^2 
\bigg]
\nn
&+C_A n_f\bigg[\frac{224}{9}-2\pi^2
 \bigg]
+C_F n_f\bigg[ 2\bigg]+n_f^2\bigg[-\frac{40}{27}
 \bigg],\,
\nn
J^g_2\big|_{\delta} &=C_A^2\bigg[\frac{20215}{162}-\frac{440}{9}\zeta_3-\frac{371}{18}\pi^2+\frac{151}{180}\pi^4 
\bigg]
\nn
&+C_A n_f \bigg[-\frac{760}{27}+\frac{8}{9}\zeta_3+\frac{109}{27}\pi^2 
\bigg]
\nn
&+C_F n_f\bigg[-\frac{55}{6}+8\zeta_3
\bigg]
+n_f^2\bigg[\frac{100}{81}-\frac{4}{27}\pi^2
\bigg],\,
\nn
J^g_3\big|_{\mathcal{D}_{5}} &=C_A^3\bigg[ 8\bigg],\,\quad
J^g_3\big|_{\mathcal{D}_{4}} =C_A^3\bigg[ -\frac{550}{9}
\bigg]+C_A^2 n_f\bigg[\frac{100}{9} \bigg],\,
\nn
J^g_3\big|_{\mathcal{D}_{3}} &=C_A^3\bigg[340-24\pi^2
\bigg]
+C_A^2 n_f\bigg[ -\frac{256}{3}
\bigg]
\nn 
&+ C_A n_f^2\bigg[ \frac{16}{3}\bigg],\,
\nn
J^g_3\big|_{\mathcal{D}_{2}} & =C_A^3\bigg[-\frac{9623}{9}+256\zeta_3+\frac{1034}{9}\pi^2 
\bigg]
\nn
&+C_A^2 n_f\bigg[\frac{3106}{9}-\frac{188}{9}\pi^2
\bigg]
+C_A C_F n_f\bigg[ 16\bigg] 
\nn
&+ C_A n_f^2\bigg[-\frac{292}{9}
\bigg] +n_f^3\bigg[\frac{8}{9} \bigg],\,
\nn
J^g_3\big|_{\mathcal{D}_{1}} &=C_A^3\bigg[ \frac{171848}{81}-\frac{7304}{9}\zeta_3-\frac{9314}{27}\pi^2+\frac{439}{45}\pi^4
\bigg] 
\nn
&+ C_A^2 n_f\bigg[-\frac{20134}{27} 
+\frac{752}{9}\zeta_3+\frac{2324}{27}\pi^2
\bigg]
\nn
&+C_A C_F n_f\bigg[ -110+64\zeta_3
\bigg]+C_A n_f^2\bigg[\frac{6652}{81}
\nn
&-\frac{16}{3}\pi^2
\bigg]
+C_F n_f^2\bigg[\frac{20}{3} \bigg] + n_f^3 \bigg[-\frac{80}{27}
 \bigg],\,
\nn
J^g_3\big|_{\mathcal{D}_{0}} &=C_A^3\bigg[ -\frac{1448021}{729}+80\zeta_5+\frac{12436}{9}\zeta_3+\frac{191963}{486}\pi^2
\nn
&-\frac{736}{9}\pi^2\zeta_3-\frac{363}{20}\pi^4
\bigg]+C_A^2 n_f\bigg[\frac{1052135}{1458}-\frac{5888}{27}\zeta_3
\nn
&-\frac{30808}{243}\pi^2+\frac{313}{90}\pi^4 
\bigg]+ C_A C_F n_f \bigg[ \frac{5599}{27}-\frac{1096}{9}\zeta_3
\nn
&-6\pi^2-\frac{8}{45}\pi^4
\bigg]+C_F^2 n_f\bigg[ -1\bigg]
+C_A n_f^2\bigg[ -\frac{116509}{1458}
\nn
&+\frac{32}{27}\zeta_3+\frac{946}{81}\pi^2
\bigg]+C_F n_f^2\bigg[-24+16\zeta_3
\bigg]
\nn
&+n_f^3\bigg[\frac{200}{81}-\frac{8}{27}\pi^2
\bigg],\,
\nn
J^g_3\big|_{\delta} &=C_A^3\bigg[\frac{55853711}{26244}-44\zeta_5-\frac{452770}{243}\zeta_3 +\frac{1600}{9}\zeta_3^2
\nn
&-\frac{2055109}{4374}\pi^2+\frac{1364}{9}\pi^2\zeta_3+\frac{53633}{1620}\pi^4-\frac{16309}{20412}\pi^6 
\bigg]
\nn
&+C_A^2 n_f\bigg[-\frac{17323633}{26244}+\frac{208}{9}\zeta_5+\frac{2734}{9}\zeta_3
\nn
&+\frac{330062}{2187}\pi^2
-\frac{88}{9}\pi^2\zeta_3-\frac{18727}{2430}\pi^4 
\bigg]
\nn
&+C_A C_F n_f\bigg[ -\frac{389369}{972}
+\frac{584}{9}\zeta_5+\frac{21200}{81}\zeta_3
\nn
&+\frac{712}{27}\pi^2-\frac{160}{9}\pi^2\zeta_3+\frac{76}{405}\pi^4
\bigg]+C_F^2 n_f\bigg[ \frac{143}{9}
\nn
&-80\zeta_5+\frac{148}{3}\zeta_3
\bigg]+C_A n_f^2\bigg[ \frac{1613639}{26244}
-\frac{1004}{243}\zeta_3
\nn
&-\frac{3656}{243}\pi^2+\frac{506}{1215}\pi^4
\bigg]+C_F n_f^2\bigg[\frac{7001}{162}-\frac{104}{3}\zeta_3
\nn
&-\frac{10}{9}\pi^2
\bigg]
+n_f^3\bigg[ -\frac{1000}{729}+\frac{40}{81}\pi^2
\bigg]\, .
\end{align}
$C_A $ and $C_F$ are the quadratic Casimir in adjoint
and fundamental  representations, respectively, and $n_f$ is the number of light flavors.

\section{III.\,CONCLUSION}
In this article, we have shown how one of the building blocks of SCET, namely, the jet function can be related to the well-known coefficient function of the DIS cross section. This novel connection provides an alternate and elegant way to obtain both quark and gluon jet functions order by order in pQCD from the known coefficient functions.   
While we confirm the three-loop quark jet function reported recently in \cite{Bruser:2018rad},
the three-loop gluon jet function presented in the article is new.
The important ingredient to obtain this result is the gluon coefficient function 
\cite{Soar:2009yh} of DIS process
up to the three-loop level in QCD,  computed using off-shell scalar particle 
scattering deep-inelastically of the massless gluons. 
We have used the factorization properties of scattering cross section and exploited universal structure
of soft and collinear dynamics to relate soft plus jet function of DIS against the jet function
in SCET.  Using quark coefficient function \cite{Vermaseren:2005qc} known up to three loops in QCD, 
we confirm the recently computed three-loop quark jet function in 
\cite{Bruser:2018rad} and present our result for the gluon jet function up to three-loop level.
The three-loop quark and gluon jet functions  
are the important ingredients to the N-jettiness IR subtraction method 
\cite{Gaunt:2015pea,Boughezal:2015dva} at N$^3$LO and to threshold resummation 
up to N$^3$LL$'$ in the SCET framework to study processes involving final state jets.
Thanks to the wealth of precise predictions in perturbative QCD for various important observables, one can 
unfold the underlying universal infrared structure QCD amplitudes and determine process-independent 
building blocks that capture infrared dynamics of high-energy scattering processes.  

\section*{Acknowledgements}
We thank T. Ahmed, A. Chakraborty, G. Das and N. Rana for useful discussions.

\bibliography{3loopjet}

\begin{thebibliography}{44}%
\makeatletter
\providecommand \@ifxundefined [1]{%
 \@ifx{#1\undefined}
}%
\providecommand \@ifnum [1]{%
 \ifnum #1\expandafter \@firstoftwo
 \else \expandafter \@secondoftwo
 \fi
}%
\providecommand \@ifx [1]{%
 \ifx #1\expandafter \@firstoftwo
 \else \expandafter \@secondoftwo
 \fi
}%
\providecommand \natexlab [1]{#1}%
\providecommand \enquote  [1]{``#1''}%
\providecommand \bibnamefont  [1]{#1}%
\providecommand \bibfnamefont [1]{#1}%
\providecommand \citenamefont [1]{#1}%
\providecommand \href@noop [0]{\@secondoftwo}%
\providecommand \href [0]{\begingroup \@sanitize@url \@href}%
\providecommand \@href[1]{\@@startlink{#1}\@@href}%
\providecommand \@@href[1]{\endgroup#1\@@endlink}%
\providecommand \@sanitize@url [0]{\catcode `\\12\catcode `\$12\catcode
  `\&12\catcode `\#12\catcode `\^12\catcode `\_12\catcode `\%12\relax}%
\providecommand \@@startlink[1]{}%
\providecommand \@@endlink[0]{}%
\providecommand \url  [0]{\begingroup\@sanitize@url \@url }%
\providecommand \@url [1]{\endgroup\@href {#1}{\urlprefix }}%
\providecommand \urlprefix  [0]{URL }%
\providecommand \Eprint [0]{\href }%
\providecommand \doibase [0]{http://dx.doi.org/}%
\providecommand \selectlanguage [0]{\@gobble}%
\providecommand \bibinfo  [0]{\@secondoftwo}%
\providecommand \bibfield  [0]{\@secondoftwo}%
\providecommand \translation [1]{[#1]}%
\providecommand \BibitemOpen [0]{}%
\providecommand \bibitemStop [0]{}%
\providecommand \bibitemNoStop [0]{.\EOS\space}%
\providecommand \EOS [0]{\spacefactor3000\relax}%
\providecommand \BibitemShut  [1]{\csname bibitem#1\endcsname}%
\let\auto@bib@innerbib\@empty
\bibitem [{\citenamefont {Vermaseren}\ \emph {et~al.}(2005)\citenamefont
  {Vermaseren}, \citenamefont {Vogt},\ and\ \citenamefont
  {Moch}}]{Vermaseren:2005qc}%
  \BibitemOpen
  \bibfield  {author} {\bibinfo {author} {\bibfnamefont {J.~A.~M.}\
  \bibnamefont {Vermaseren}}, \bibinfo {author} {\bibfnamefont
  {A.}~\bibnamefont {Vogt}}, \ and\ \bibinfo {author} {\bibfnamefont
  {S.}~\bibnamefont {Moch}},\ }\href {\doibase 10.1016/j.nuclphysb.2005.06.020}
  {\bibfield  {journal} {\bibinfo  {journal} {Nucl. Phys.}\ }\textbf {\bibinfo
  {volume} {B724}},\ \bibinfo {pages} {3} (\bibinfo {year} {2005})},\ \Eprint
  {http://arxiv.org/abs/hep-ph/0504242} {arXiv:hep-ph/0504242 [hep-ph]}
  \BibitemShut {NoStop}%
\bibitem [{\citenamefont {Soar}\ \emph {et~al.}(2010)\citenamefont {Soar},
  \citenamefont {Moch}, \citenamefont {Vermaseren},\ and\ \citenamefont
  {Vogt}}]{Soar:2009yh}%
  \BibitemOpen
  \bibfield  {author} {\bibinfo {author} {\bibfnamefont {G.}~\bibnamefont
  {Soar}}, \bibinfo {author} {\bibfnamefont {S.}~\bibnamefont {Moch}}, \bibinfo
  {author} {\bibfnamefont {J.~A.~M.}\ \bibnamefont {Vermaseren}}, \ and\
  \bibinfo {author} {\bibfnamefont {A.}~\bibnamefont {Vogt}},\ }\href {\doibase
  10.1016/j.nuclphysb.2010.02.003} {\bibfield  {journal} {\bibinfo  {journal}
  {Nucl. Phys.}\ }\textbf {\bibinfo {volume} {B832}},\ \bibinfo {pages} {152}
  (\bibinfo {year} {2010})},\ \Eprint {http://arxiv.org/abs/0912.0369}
  {arXiv:0912.0369 [hep-ph]} \BibitemShut {NoStop}%
\bibitem [{\citenamefont {{Br\"user}}\ \emph {et~al.}(2018)\citenamefont
  {{Br\"user}}, \citenamefont {Liu},\ and\ \citenamefont
  {Stahlhofen}}]{Bruser:2018rad}%
  \BibitemOpen
  \bibfield  {author} {\bibinfo {author} {\bibfnamefont {R.}~\bibnamefont
  {{Br\"user}}}, \bibinfo {author} {\bibfnamefont {Z.~L.}\ \bibnamefont {Liu}},
  \ and\ \bibinfo {author} {\bibfnamefont {M.}~\bibnamefont {Stahlhofen}},\
  }\href {\doibase 10.1103/PhysRevLett.121.072003} {\bibfield  {journal}
  {\bibinfo  {journal} {Phys. Rev. Lett.}\ }\textbf {\bibinfo {volume} {121}},\
  \bibinfo {pages} {072003} (\bibinfo {year} {2018})},\ \Eprint
  {http://arxiv.org/abs/1804.09722} {arXiv:1804.09722 [hep-ph]} \BibitemShut
  {NoStop}%
\bibitem [{\citenamefont {Bauer}\ \emph {et~al.}(2000)\citenamefont {Bauer},
  \citenamefont {Fleming},\ and\ \citenamefont {Luke}}]{Bauer:2000ew}%
  \BibitemOpen
  \bibfield  {author} {\bibinfo {author} {\bibfnamefont {C.~W.}\ \bibnamefont
  {Bauer}}, \bibinfo {author} {\bibfnamefont {S.}~\bibnamefont {Fleming}}, \
  and\ \bibinfo {author} {\bibfnamefont {M.~E.}\ \bibnamefont {Luke}},\ }\href
  {\doibase 10.1103/PhysRevD.63.014006} {\bibfield  {journal} {\bibinfo
  {journal} {Phys. Rev.}\ }\textbf {\bibinfo {volume} {D63}},\ \bibinfo {pages}
  {014006} (\bibinfo {year} {2000})},\ \Eprint
  {http://arxiv.org/abs/hep-ph/0005275} {arXiv:hep-ph/0005275 [hep-ph]}
  \BibitemShut {NoStop}%
\bibitem [{\citenamefont {Bauer}\ \emph {et~al.}(2001)\citenamefont {Bauer},
  \citenamefont {Fleming}, \citenamefont {Pirjol},\ and\ \citenamefont
  {Stewart}}]{Bauer:2000yr}%
  \BibitemOpen
  \bibfield  {author} {\bibinfo {author} {\bibfnamefont {C.~W.}\ \bibnamefont
  {Bauer}}, \bibinfo {author} {\bibfnamefont {S.}~\bibnamefont {Fleming}},
  \bibinfo {author} {\bibfnamefont {D.}~\bibnamefont {Pirjol}}, \ and\ \bibinfo
  {author} {\bibfnamefont {I.~W.}\ \bibnamefont {Stewart}},\ }\href {\doibase
  10.1103/PhysRevD.63.114020} {\bibfield  {journal} {\bibinfo  {journal} {Phys.
  Rev.}\ }\textbf {\bibinfo {volume} {D63}},\ \bibinfo {pages} {114020}
  (\bibinfo {year} {2001})},\ \Eprint {http://arxiv.org/abs/hep-ph/0011336}
  {arXiv:hep-ph/0011336 [hep-ph]} \BibitemShut {NoStop}%
\bibitem [{\citenamefont {Bauer}\ and\ \citenamefont
  {Stewart}(2001)}]{Bauer:2001ct}%
  \BibitemOpen
  \bibfield  {author} {\bibinfo {author} {\bibfnamefont {C.~W.}\ \bibnamefont
  {Bauer}}\ and\ \bibinfo {author} {\bibfnamefont {I.~W.}\ \bibnamefont
  {Stewart}},\ }\href {\doibase 10.1016/S0370-2693(01)00902-9} {\bibfield
  {journal} {\bibinfo  {journal} {Phys. Lett.}\ }\textbf {\bibinfo {volume}
  {B516}},\ \bibinfo {pages} {134} (\bibinfo {year} {2001})},\ \Eprint
  {http://arxiv.org/abs/hep-ph/0107001} {arXiv:hep-ph/0107001 [hep-ph]}
  \BibitemShut {NoStop}%
\bibitem [{\citenamefont {Bauer}\ \emph
  {et~al.}(2002{\natexlab{a}})\citenamefont {Bauer}, \citenamefont {Pirjol},\
  and\ \citenamefont {Stewart}}]{Bauer:2001yt}%
  \BibitemOpen
  \bibfield  {author} {\bibinfo {author} {\bibfnamefont {C.~W.}\ \bibnamefont
  {Bauer}}, \bibinfo {author} {\bibfnamefont {D.}~\bibnamefont {Pirjol}}, \
  and\ \bibinfo {author} {\bibfnamefont {I.~W.}\ \bibnamefont {Stewart}},\
  }\href {\doibase 10.1103/PhysRevD.65.054022} {\bibfield  {journal} {\bibinfo
  {journal} {Phys. Rev.}\ }\textbf {\bibinfo {volume} {D65}},\ \bibinfo {pages}
  {054022} (\bibinfo {year} {2002}{\natexlab{a}})},\ \Eprint
  {http://arxiv.org/abs/hep-ph/0109045} {arXiv:hep-ph/0109045 [hep-ph]}
  \BibitemShut {NoStop}%
\bibitem [{\citenamefont {Bauer}\ \emph
  {et~al.}(2002{\natexlab{b}})\citenamefont {Bauer}, \citenamefont {Fleming},
  \citenamefont {Pirjol}, \citenamefont {Rothstein},\ and\ \citenamefont
  {Stewart}}]{Bauer:2002nz}%
  \BibitemOpen
  \bibfield  {author} {\bibinfo {author} {\bibfnamefont {C.~W.}\ \bibnamefont
  {Bauer}}, \bibinfo {author} {\bibfnamefont {S.}~\bibnamefont {Fleming}},
  \bibinfo {author} {\bibfnamefont {D.}~\bibnamefont {Pirjol}}, \bibinfo
  {author} {\bibfnamefont {I.~Z.}\ \bibnamefont {Rothstein}}, \ and\ \bibinfo
  {author} {\bibfnamefont {I.~W.}\ \bibnamefont {Stewart}},\ }\href {\doibase
  10.1103/PhysRevD.66.014017} {\bibfield  {journal} {\bibinfo  {journal} {Phys.
  Rev.}\ }\textbf {\bibinfo {volume} {D66}},\ \bibinfo {pages} {014017}
  (\bibinfo {year} {2002}{\natexlab{b}})},\ \Eprint
  {http://arxiv.org/abs/hep-ph/0202088} {arXiv:hep-ph/0202088 [hep-ph]}
  \BibitemShut {NoStop}%
\bibitem [{\citenamefont {Beneke}\ \emph {et~al.}(2002)\citenamefont {Beneke},
  \citenamefont {Chapovsky}, \citenamefont {Diehl},\ and\ \citenamefont
  {Feldmann}}]{Beneke:2002ph}%
  \BibitemOpen
  \bibfield  {author} {\bibinfo {author} {\bibfnamefont {M.}~\bibnamefont
  {Beneke}}, \bibinfo {author} {\bibfnamefont {A.~P.}\ \bibnamefont
  {Chapovsky}}, \bibinfo {author} {\bibfnamefont {M.}~\bibnamefont {Diehl}}, \
  and\ \bibinfo {author} {\bibfnamefont {T.}~\bibnamefont {Feldmann}},\ }\href
  {\doibase 10.1016/S0550-3213(02)00687-9} {\bibfield  {journal} {\bibinfo
  {journal} {Nucl. Phys.}\ }\textbf {\bibinfo {volume} {B643}},\ \bibinfo
  {pages} {431} (\bibinfo {year} {2002})},\ \Eprint
  {http://arxiv.org/abs/hep-ph/0206152} {arXiv:hep-ph/0206152 [hep-ph]}
  \BibitemShut {NoStop}%
\bibitem [{\citenamefont {Li}\ \emph {et~al.}(2014)\citenamefont {Li},
  \citenamefont {von Manteuffel}, \citenamefont {Schabinger},\ and\
  \citenamefont {Zhu}}]{Li:2014bfa}%
  \BibitemOpen
  \bibfield  {author} {\bibinfo {author} {\bibfnamefont {Y.}~\bibnamefont
  {Li}}, \bibinfo {author} {\bibfnamefont {A.}~\bibnamefont {von Manteuffel}},
  \bibinfo {author} {\bibfnamefont {R.~M.}\ \bibnamefont {Schabinger}}, \ and\
  \bibinfo {author} {\bibfnamefont {H.~X.}\ \bibnamefont {Zhu}},\ }\href
  {\doibase 10.1103/PhysRevD.90.053006} {\bibfield  {journal} {\bibinfo
  {journal} {Phys. Rev.}\ }\textbf {\bibinfo {volume} {D90}},\ \bibinfo {pages}
  {053006} (\bibinfo {year} {2014})},\ \Eprint {http://arxiv.org/abs/1404.5839}
  {arXiv:1404.5839 [hep-ph]} \BibitemShut {NoStop}%
\bibitem [{\citenamefont {Bauer}\ and\ \citenamefont
  {Manohar}(2004)}]{Bauer:2003pi}%
  \BibitemOpen
  \bibfield  {author} {\bibinfo {author} {\bibfnamefont {C.~W.}\ \bibnamefont
  {Bauer}}\ and\ \bibinfo {author} {\bibfnamefont {A.~V.}\ \bibnamefont
  {Manohar}},\ }\href {\doibase 10.1103/PhysRevD.70.034024} {\bibfield
  {journal} {\bibinfo  {journal} {Phys. Rev.}\ }\textbf {\bibinfo {volume}
  {D70}},\ \bibinfo {pages} {034024} (\bibinfo {year} {2004})},\ \Eprint
  {http://arxiv.org/abs/hep-ph/0312109} {arXiv:hep-ph/0312109 [hep-ph]}
  \BibitemShut {NoStop}%
\bibitem [{\citenamefont {Bosch}\ \emph {et~al.}(2004)\citenamefont {Bosch},
  \citenamefont {Lange}, \citenamefont {Neubert},\ and\ \citenamefont
  {Paz}}]{Bosch:2004th}%
  \BibitemOpen
  \bibfield  {author} {\bibinfo {author} {\bibfnamefont {S.~W.}\ \bibnamefont
  {Bosch}}, \bibinfo {author} {\bibfnamefont {B.~O.}\ \bibnamefont {Lange}},
  \bibinfo {author} {\bibfnamefont {M.}~\bibnamefont {Neubert}}, \ and\
  \bibinfo {author} {\bibfnamefont {G.}~\bibnamefont {Paz}},\ }\href {\doibase
  10.1016/j.nuclphysb.2004.07.041} {\bibfield  {journal} {\bibinfo  {journal}
  {Nucl. Phys.}\ }\textbf {\bibinfo {volume} {B699}},\ \bibinfo {pages} {335}
  (\bibinfo {year} {2004})},\ \Eprint {http://arxiv.org/abs/hep-ph/0402094}
  {arXiv:hep-ph/0402094 [hep-ph]} \BibitemShut {NoStop}%
\bibitem [{\citenamefont {Becher}\ and\ \citenamefont
  {Neubert}(2006)}]{Becher:2006qw}%
  \BibitemOpen
  \bibfield  {author} {\bibinfo {author} {\bibfnamefont {T.}~\bibnamefont
  {Becher}}\ and\ \bibinfo {author} {\bibfnamefont {M.}~\bibnamefont
  {Neubert}},\ }\href {\doibase 10.1016/j.physletb.2006.04.046} {\bibfield
  {journal} {\bibinfo  {journal} {Phys. Lett.}\ }\textbf {\bibinfo {volume}
  {B637}},\ \bibinfo {pages} {251} (\bibinfo {year} {2006})},\ \Eprint
  {http://arxiv.org/abs/hep-ph/0603140} {arXiv:hep-ph/0603140 [hep-ph]}
  \BibitemShut {NoStop}%
\bibitem [{\citenamefont {Becher}\ and\ \citenamefont
  {Schwartz}(2010)}]{Becher:2009th}%
  \BibitemOpen
  \bibfield  {author} {\bibinfo {author} {\bibfnamefont {T.}~\bibnamefont
  {Becher}}\ and\ \bibinfo {author} {\bibfnamefont {M.~D.}\ \bibnamefont
  {Schwartz}},\ }\href {\doibase 10.1007/JHEP02(2010)040} {\bibfield  {journal}
  {\bibinfo  {journal} {JHEP}\ }\textbf {\bibinfo {volume} {02}},\ \bibinfo
  {pages} {040} (\bibinfo {year} {2010})},\ \Eprint
  {http://arxiv.org/abs/0911.0681} {arXiv:0911.0681 [hep-ph]} \BibitemShut
  {NoStop}%
\bibitem [{\citenamefont {Becher}\ and\ \citenamefont
  {Bell}(2011)}]{Becher:2010pd}%
  \BibitemOpen
  \bibfield  {author} {\bibinfo {author} {\bibfnamefont {T.}~\bibnamefont
  {Becher}}\ and\ \bibinfo {author} {\bibfnamefont {G.}~\bibnamefont {Bell}},\
  }\href {\doibase 10.1016/j.physletb.2010.11.036} {\bibfield  {journal}
  {\bibinfo  {journal} {Phys. Lett.}\ }\textbf {\bibinfo {volume} {B695}},\
  \bibinfo {pages} {252} (\bibinfo {year} {2011})},\ \Eprint
  {http://arxiv.org/abs/1008.1936} {arXiv:1008.1936 [hep-ph]} \BibitemShut
  {NoStop}%
\bibitem [{\citenamefont {Ravindran}(2006{\natexlab{a}})}]{Ravindran:2005vv}%
  \BibitemOpen
  \bibfield  {author} {\bibinfo {author} {\bibfnamefont {V.}~\bibnamefont
  {Ravindran}},\ }\href {\doibase 10.1016/j.nuclphysb.2006.04.008} {\bibfield
  {journal} {\bibinfo  {journal} {Nucl. Phys.}\ }\textbf {\bibinfo {volume}
  {B746}},\ \bibinfo {pages} {58} (\bibinfo {year} {2006}{\natexlab{a}})},\
  \Eprint {http://arxiv.org/abs/hep-ph/0512249} {arXiv:hep-ph/0512249 [hep-ph]}
  \BibitemShut {NoStop}%
\bibitem [{\citenamefont {Ravindran}(2006{\natexlab{b}})}]{Ravindran:2006cg}%
  \BibitemOpen
  \bibfield  {author} {\bibinfo {author} {\bibfnamefont {V.}~\bibnamefont
  {Ravindran}},\ }\href {\doibase 10.1016/j.nuclphysb.2006.06.025} {\bibfield
  {journal} {\bibinfo  {journal} {Nucl. Phys.}\ }\textbf {\bibinfo {volume}
  {B752}},\ \bibinfo {pages} {173} (\bibinfo {year} {2006}{\natexlab{b}})},\
  \Eprint {http://arxiv.org/abs/hep-ph/0603041} {arXiv:hep-ph/0603041 [hep-ph]}
  \BibitemShut {NoStop}%
\bibitem [{\citenamefont {Ravindran}\ \emph {et~al.}(2007)\citenamefont
  {Ravindran}, \citenamefont {Smith},\ and\ \citenamefont {van
  Neerven}}]{Ravindran:2006bu}%
  \BibitemOpen
  \bibfield  {author} {\bibinfo {author} {\bibfnamefont {V.}~\bibnamefont
  {Ravindran}}, \bibinfo {author} {\bibfnamefont {J.}~\bibnamefont {Smith}}, \
  and\ \bibinfo {author} {\bibfnamefont {W.~L.}\ \bibnamefont {van Neerven}},\
  }\href {\doibase 10.1016/j.nuclphysb.2007.01.005} {\bibfield  {journal}
  {\bibinfo  {journal} {Nucl. Phys.}\ }\textbf {\bibinfo {volume} {B767}},\
  \bibinfo {pages} {100} (\bibinfo {year} {2007})},\ \Eprint
  {http://arxiv.org/abs/hep-ph/0608308} {arXiv:hep-ph/0608308 [hep-ph]}
  \BibitemShut {NoStop}%
\bibitem [{\citenamefont {Banerjee}\ \emph
  {et~al.}(2018{\natexlab{a}})\citenamefont {Banerjee}, \citenamefont {Das},
  \citenamefont {Dhani},\ and\ \citenamefont {Ravindran}}]{Banerjee:2017cfc}%
  \BibitemOpen
  \bibfield  {author} {\bibinfo {author} {\bibfnamefont {P.}~\bibnamefont
  {Banerjee}}, \bibinfo {author} {\bibfnamefont {G.}~\bibnamefont {Das}},
  \bibinfo {author} {\bibfnamefont {P.~K.}\ \bibnamefont {Dhani}}, \ and\
  \bibinfo {author} {\bibfnamefont {V.}~\bibnamefont {Ravindran}},\ }\href
  {\doibase 10.1103/PhysRevD.97.054024} {\bibfield  {journal} {\bibinfo
  {journal} {Phys. Rev.}\ }\textbf {\bibinfo {volume} {D97}},\ \bibinfo {pages}
  {054024} (\bibinfo {year} {2018}{\natexlab{a}})},\ \Eprint
  {http://arxiv.org/abs/1708.05706} {arXiv:1708.05706 [hep-ph]} \BibitemShut
  {NoStop}%
\bibitem [{\citenamefont {Banerjee}\ \emph
  {et~al.}(2018{\natexlab{b}})\citenamefont {Banerjee}, \citenamefont {Das},
  \citenamefont {Dhani},\ and\ \citenamefont {Ravindran}}]{Banerjee:2018vvb}%
  \BibitemOpen
  \bibfield  {author} {\bibinfo {author} {\bibfnamefont {P.}~\bibnamefont
  {Banerjee}}, \bibinfo {author} {\bibfnamefont {G.}~\bibnamefont {Das}},
  \bibinfo {author} {\bibfnamefont {P.~K.}\ \bibnamefont {Dhani}}, \ and\
  \bibinfo {author} {\bibfnamefont {V.}~\bibnamefont {Ravindran}},\ }\href
  {\doibase 10.1103/PhysRevD.98.054018} {\bibfield  {journal} {\bibinfo
  {journal} {Phys. Rev.}\ }\textbf {\bibinfo {volume} {D98}},\ \bibinfo {pages}
  {054018} (\bibinfo {year} {2018}{\natexlab{b}})},\ \Eprint
  {http://arxiv.org/abs/1805.01186} {arXiv:1805.01186 [hep-ph]} \BibitemShut
  {NoStop}%
\bibitem [{\citenamefont {Chetyrkin}\ \emph {et~al.}(1998)\citenamefont
  {Chetyrkin}, \citenamefont {Kniehl},\ and\ \citenamefont
  {Steinhauser}}]{Chetyrkin:1997un}%
  \BibitemOpen
  \bibfield  {author} {\bibinfo {author} {\bibfnamefont {K.~G.}\ \bibnamefont
  {Chetyrkin}}, \bibinfo {author} {\bibfnamefont {B.~A.}\ \bibnamefont
  {Kniehl}}, \ and\ \bibinfo {author} {\bibfnamefont {M.}~\bibnamefont
  {Steinhauser}},\ }\href {\doibase 10.1016/S0550-3213(98)81004-3,
  10.1016/S0550-3213(97)00649-4} {\bibfield  {journal} {\bibinfo  {journal}
  {Nucl. Phys.}\ }\textbf {\bibinfo {volume} {B510}},\ \bibinfo {pages} {61}
  (\bibinfo {year} {1998})},\ \Eprint {http://arxiv.org/abs/hep-ph/9708255}
  {arXiv:hep-ph/9708255 [hep-ph]} \BibitemShut {NoStop}%
\bibitem [{\citenamefont {Moch}\ \emph {et~al.}(2004)\citenamefont {Moch},
  \citenamefont {Vermaseren},\ and\ \citenamefont {Vogt}}]{Moch:2004pa}%
  \BibitemOpen
  \bibfield  {author} {\bibinfo {author} {\bibfnamefont {S.}~\bibnamefont
  {Moch}}, \bibinfo {author} {\bibfnamefont {J.~A.~M.}\ \bibnamefont
  {Vermaseren}}, \ and\ \bibinfo {author} {\bibfnamefont {A.}~\bibnamefont
  {Vogt}},\ }\href {\doibase 10.1016/j.nuclphysb.2004.03.030} {\bibfield
  {journal} {\bibinfo  {journal} {Nucl. Phys.}\ }\textbf {\bibinfo {volume}
  {B688}},\ \bibinfo {pages} {101} (\bibinfo {year} {2004})},\ \Eprint
  {http://arxiv.org/abs/hep-ph/0403192} {arXiv:hep-ph/0403192 [hep-ph]}
  \BibitemShut {NoStop}%
\bibitem [{\citenamefont {Vogt}\ \emph {et~al.}(2004)\citenamefont {Vogt},
  \citenamefont {Moch},\ and\ \citenamefont {Vermaseren}}]{Vogt:2004mw}%
  \BibitemOpen
  \bibfield  {author} {\bibinfo {author} {\bibfnamefont {A.}~\bibnamefont
  {Vogt}}, \bibinfo {author} {\bibfnamefont {S.}~\bibnamefont {Moch}}, \ and\
  \bibinfo {author} {\bibfnamefont {J.~A.~M.}\ \bibnamefont {Vermaseren}},\
  }\href {\doibase 10.1016/j.nuclphysb.2004.04.024} {\bibfield  {journal}
  {\bibinfo  {journal} {Nucl. Phys.}\ }\textbf {\bibinfo {volume} {B691}},\
  \bibinfo {pages} {129} (\bibinfo {year} {2004})},\ \Eprint
  {http://arxiv.org/abs/hep-ph/0404111} {arXiv:hep-ph/0404111 [hep-ph]}
  \BibitemShut {NoStop}%
\bibitem [{\citenamefont {Davies}\ \emph {et~al.}(2017)\citenamefont {Davies},
  \citenamefont {Vogt}, \citenamefont {Ruijl}, \citenamefont {Ueda},\ and\
  \citenamefont {Vermaseren}}]{Davies:2016jie}%
  \BibitemOpen
  \bibfield  {author} {\bibinfo {author} {\bibfnamefont {J.}~\bibnamefont
  {Davies}}, \bibinfo {author} {\bibfnamefont {A.}~\bibnamefont {Vogt}},
  \bibinfo {author} {\bibfnamefont {B.}~\bibnamefont {Ruijl}}, \bibinfo
  {author} {\bibfnamefont {T.}~\bibnamefont {Ueda}}, \ and\ \bibinfo {author}
  {\bibfnamefont {J.~A.~M.}\ \bibnamefont {Vermaseren}},\ }\href {\doibase
  10.1016/j.nuclphysb.2016.12.012} {\bibfield  {journal} {\bibinfo  {journal}
  {Nucl. Phys.}\ }\textbf {\bibinfo {volume} {B915}},\ \bibinfo {pages} {335}
  (\bibinfo {year} {2017})},\ \Eprint {http://arxiv.org/abs/1610.07477}
  {arXiv:1610.07477 [hep-ph]} \BibitemShut {NoStop}%
\bibitem [{\citenamefont {Sudakov}(1956)}]{Sudakov:1954sw}%
  \BibitemOpen
  \bibfield  {author} {\bibinfo {author} {\bibfnamefont {V.~V.}\ \bibnamefont
  {Sudakov}},\ }\href@noop {} {\bibfield  {journal} {\bibinfo  {journal} {Sov.
  Phys. JETP}\ }\textbf {\bibinfo {volume} {3}},\ \bibinfo {pages} {65}
  (\bibinfo {year} {1956})},\ \bibinfo {note} {[Zh. Eksp. Teor.
  Fiz.30,87(1956)]}\BibitemShut {NoStop}%
\bibitem [{\citenamefont {Mueller}(1979)}]{Mueller:1979ih}%
  \BibitemOpen
  \bibfield  {author} {\bibinfo {author} {\bibfnamefont {A.~H.}\ \bibnamefont
  {Mueller}},\ }\href {\doibase 10.1103/PhysRevD.20.2037} {\bibfield  {journal}
  {\bibinfo  {journal} {Phys. Rev.}\ }\textbf {\bibinfo {volume} {D20}},\
  \bibinfo {pages} {2037} (\bibinfo {year} {1979})}\BibitemShut {NoStop}%
\bibitem [{\citenamefont {Collins}(1980)}]{Collins:1980ih}%
  \BibitemOpen
  \bibfield  {author} {\bibinfo {author} {\bibfnamefont {J.~C.}\ \bibnamefont
  {Collins}},\ }\href {\doibase 10.1103/PhysRevD.22.1478} {\bibfield  {journal}
  {\bibinfo  {journal} {Phys. Rev.}\ }\textbf {\bibinfo {volume} {D22}},\
  \bibinfo {pages} {1478} (\bibinfo {year} {1980})}\BibitemShut {NoStop}%
\bibitem [{\citenamefont {Sen}(1981)}]{Sen:1981sd}%
  \BibitemOpen
  \bibfield  {author} {\bibinfo {author} {\bibfnamefont {A.}~\bibnamefont
  {Sen}},\ }\href {\doibase 10.1103/PhysRevD.24.3281} {\bibfield  {journal}
  {\bibinfo  {journal} {Phys. Rev.}\ }\textbf {\bibinfo {volume} {D24}},\
  \bibinfo {pages} {3281} (\bibinfo {year} {1981})}\BibitemShut {NoStop}%
\bibitem [{\citenamefont {Moch}\ \emph {et~al.}(2005)\citenamefont {Moch},
  \citenamefont {Vermaseren},\ and\ \citenamefont {Vogt}}]{Moch:2005id}%
  \BibitemOpen
  \bibfield  {author} {\bibinfo {author} {\bibfnamefont {S.}~\bibnamefont
  {Moch}}, \bibinfo {author} {\bibfnamefont {J.~A.~M.}\ \bibnamefont
  {Vermaseren}}, \ and\ \bibinfo {author} {\bibfnamefont {A.}~\bibnamefont
  {Vogt}},\ }\href {\doibase 10.1088/1126-6708/2005/08/049} {\bibfield
  {journal} {\bibinfo  {journal} {JHEP}\ }\textbf {\bibinfo {volume} {08}},\
  \bibinfo {pages} {049} (\bibinfo {year} {2005})},\ \Eprint
  {http://arxiv.org/abs/hep-ph/0507039} {arXiv:hep-ph/0507039 [hep-ph]}
  \BibitemShut {NoStop}%
\bibitem [{\citenamefont {Catani}\ and\ \citenamefont
  {Trentadue}(1989)}]{Catani:1989ne}%
  \BibitemOpen
  \bibfield  {author} {\bibinfo {author} {\bibfnamefont {S.}~\bibnamefont
  {Catani}}\ and\ \bibinfo {author} {\bibfnamefont {L.}~\bibnamefont
  {Trentadue}},\ }\href {\doibase 10.1016/0550-3213(89)90273-3} {\bibfield
  {journal} {\bibinfo  {journal} {Nucl. Phys.}\ }\textbf {\bibinfo {volume}
  {B327}},\ \bibinfo {pages} {323} (\bibinfo {year} {1989})}\BibitemShut
  {NoStop}%
\bibitem [{\citenamefont {Catani}\ and\ \citenamefont
  {Trentadue}(1991)}]{Catani:1990rp}%
  \BibitemOpen
  \bibfield  {author} {\bibinfo {author} {\bibfnamefont {S.}~\bibnamefont
  {Catani}}\ and\ \bibinfo {author} {\bibfnamefont {L.}~\bibnamefont
  {Trentadue}},\ }\href {\doibase 10.1016/0550-3213(91)90506-S} {\bibfield
  {journal} {\bibinfo  {journal} {Nucl. Phys.}\ }\textbf {\bibinfo {volume}
  {B353}},\ \bibinfo {pages} {183} (\bibinfo {year} {1991})}\BibitemShut
  {NoStop}%
\bibitem [{\citenamefont {Vogt}(2001)}]{Vogt:2000ci}%
  \BibitemOpen
  \bibfield  {author} {\bibinfo {author} {\bibfnamefont {A.}~\bibnamefont
  {Vogt}},\ }\href {\doibase 10.1016/S0370-2693(00)01344-7} {\bibfield
  {journal} {\bibinfo  {journal} {Phys. Lett.}\ }\textbf {\bibinfo {volume}
  {B497}},\ \bibinfo {pages} {228} (\bibinfo {year} {2001})},\ \Eprint
  {http://arxiv.org/abs/hep-ph/0010146} {arXiv:hep-ph/0010146 [hep-ph]}
  \BibitemShut {NoStop}%
\bibitem [{\citenamefont {Ravindran}\ \emph {et~al.}(2005)\citenamefont
  {Ravindran}, \citenamefont {Smith},\ and\ \citenamefont {van
  Neerven}}]{Ravindran:2004mb}%
  \BibitemOpen
  \bibfield  {author} {\bibinfo {author} {\bibfnamefont {V.}~\bibnamefont
  {Ravindran}}, \bibinfo {author} {\bibfnamefont {J.}~\bibnamefont {Smith}}, \
  and\ \bibinfo {author} {\bibfnamefont {W.~L.}\ \bibnamefont {van Neerven}},\
  }\href {\doibase 10.1016/j.nuclphysb.2004.10.039} {\bibfield  {journal}
  {\bibinfo  {journal} {Nucl. Phys.}\ }\textbf {\bibinfo {volume} {B704}},\
  \bibinfo {pages} {332} (\bibinfo {year} {2005})},\ \Eprint
  {http://arxiv.org/abs/hep-ph/0408315} {arXiv:hep-ph/0408315 [hep-ph]}
  \BibitemShut {NoStop}%
\bibitem [{\citenamefont {Tarasov}\ \emph {et~al.}(1980)\citenamefont
  {Tarasov}, \citenamefont {Vladimirov},\ and\ \citenamefont
  {Zharkov}}]{Tarasov:1980au}%
  \BibitemOpen
  \bibfield  {author} {\bibinfo {author} {\bibfnamefont {O.~V.}\ \bibnamefont
  {Tarasov}}, \bibinfo {author} {\bibfnamefont {A.~A.}\ \bibnamefont
  {Vladimirov}}, \ and\ \bibinfo {author} {\bibfnamefont {A.~{\relax Yu}.}\
  \bibnamefont {Zharkov}},\ }\href {\doibase 10.1016/0370-2693(80)90358-5}
  {\bibfield  {journal} {\bibinfo  {journal} {Phys. Lett.}\ }\textbf {\bibinfo
  {volume} {B93}},\ \bibinfo {pages} {429} (\bibinfo {year}
  {1980})}\BibitemShut {NoStop}%
\bibitem [{\citenamefont {van Neerven}\ and\ \citenamefont
  {Zijlstra}(1991)}]{vanNeerven:1991nn}%
  \BibitemOpen
  \bibfield  {author} {\bibinfo {author} {\bibfnamefont {W.~L.}\ \bibnamefont
  {van Neerven}}\ and\ \bibinfo {author} {\bibfnamefont {E.~B.}\ \bibnamefont
  {Zijlstra}},\ }\href {\doibase 10.1016/0370-2693(91)91024-P} {\bibfield
  {journal} {\bibinfo  {journal} {Phys. Lett.}\ }\textbf {\bibinfo {volume}
  {B272}},\ \bibinfo {pages} {127} (\bibinfo {year} {1991})}\BibitemShut
  {NoStop}%
\bibitem [{\citenamefont {Zijlstra}\ and\ \citenamefont {van
  Neerven}(1992{\natexlab{a}})}]{Zijlstra:1992kj}%
  \BibitemOpen
  \bibfield  {author} {\bibinfo {author} {\bibfnamefont {E.~B.}\ \bibnamefont
  {Zijlstra}}\ and\ \bibinfo {author} {\bibfnamefont {W.~L.}\ \bibnamefont {van
  Neerven}},\ }\href {\doibase 10.1016/0370-2693(92)91277-G} {\bibfield
  {journal} {\bibinfo  {journal} {Phys. Lett.}\ }\textbf {\bibinfo {volume}
  {B297}},\ \bibinfo {pages} {377} (\bibinfo {year}
  {1992}{\natexlab{a}})}\BibitemShut {NoStop}%
\bibitem [{\citenamefont {Zijlstra}\ and\ \citenamefont {van
  Neerven}(1991)}]{Zijlstra:1991qc}%
  \BibitemOpen
  \bibfield  {author} {\bibinfo {author} {\bibfnamefont {E.~B.}\ \bibnamefont
  {Zijlstra}}\ and\ \bibinfo {author} {\bibfnamefont {W.~L.}\ \bibnamefont {van
  Neerven}},\ }\href {\doibase 10.1016/0370-2693(91)90301-6} {\bibfield
  {journal} {\bibinfo  {journal} {Phys. Lett.}\ }\textbf {\bibinfo {volume}
  {B273}},\ \bibinfo {pages} {476} (\bibinfo {year} {1991})}\BibitemShut
  {NoStop}%
\bibitem [{\citenamefont {Zijlstra}\ and\ \citenamefont {van
  Neerven}(1992{\natexlab{b}})}]{Zijlstra:1992qd}%
  \BibitemOpen
  \bibfield  {author} {\bibinfo {author} {\bibfnamefont {E.~B.}\ \bibnamefont
  {Zijlstra}}\ and\ \bibinfo {author} {\bibfnamefont {W.~L.}\ \bibnamefont {van
  Neerven}},\ }\href {\doibase 10.1016/0550-3213(92)90087-R} {\bibfield
  {journal} {\bibinfo  {journal} {Nucl. Phys.}\ }\textbf {\bibinfo {volume}
  {B383}},\ \bibinfo {pages} {525} (\bibinfo {year}
  {1992}{\natexlab{b}})}\BibitemShut {NoStop}%
\bibitem [{\citenamefont {Hamberg}\ \emph {et~al.}(1991)\citenamefont
  {Hamberg}, \citenamefont {van Neerven},\ and\ \citenamefont
  {Matsuura}}]{Hamberg:1990np}%
  \BibitemOpen
  \bibfield  {author} {\bibinfo {author} {\bibfnamefont {R.}~\bibnamefont
  {Hamberg}}, \bibinfo {author} {\bibfnamefont {W.~L.}\ \bibnamefont {van
  Neerven}}, \ and\ \bibinfo {author} {\bibfnamefont {T.}~\bibnamefont
  {Matsuura}},\ }\href {\doibase 10.1016/S0550-3213(02)00814-3,
  10.1016/0550-3213(91)90064-5} {\bibfield  {journal} {\bibinfo  {journal}
  {Nucl. Phys.}\ }\textbf {\bibinfo {volume} {B359}},\ \bibinfo {pages} {343}
  (\bibinfo {year} {1991})},\ \bibinfo {note} {[Erratum: Nucl.
  Phys.B644,403(2002)]}\BibitemShut {NoStop}%
\bibitem [{\citenamefont {Sterman}\ and\ \citenamefont
  {Weinberg}(1977)}]{Sterman:1977wj}%
  \BibitemOpen
  \bibfield  {author} {\bibinfo {author} {\bibfnamefont {G.~F.}\ \bibnamefont
  {Sterman}}\ and\ \bibinfo {author} {\bibfnamefont {S.}~\bibnamefont
  {Weinberg}},\ }\href {\doibase 10.1103/PhysRevLett.39.1436} {\bibfield
  {journal} {\bibinfo  {journal} {Phys. Rev. Lett.}\ }\textbf {\bibinfo
  {volume} {39}},\ \bibinfo {pages} {1436} (\bibinfo {year}
  {1977})}\BibitemShut {NoStop}%
\bibitem [{\citenamefont {Salam}\ and\ \citenamefont
  {Soyez}(2007)}]{Salam:2007xv}%
  \BibitemOpen
  \bibfield  {author} {\bibinfo {author} {\bibfnamefont {G.~P.}\ \bibnamefont
  {Salam}}\ and\ \bibinfo {author} {\bibfnamefont {G.}~\bibnamefont {Soyez}},\
  }\href {\doibase 10.1088/1126-6708/2007/05/086} {\bibfield  {journal}
  {\bibinfo  {journal} {JHEP}\ }\textbf {\bibinfo {volume} {05}},\ \bibinfo
  {pages} {086} (\bibinfo {year} {2007})},\ \Eprint
  {http://arxiv.org/abs/0704.0292} {arXiv:0704.0292 [hep-ph]} \BibitemShut
  {NoStop}%
\bibitem [{\citenamefont {Cacciari}\ \emph {et~al.}(2008)\citenamefont
  {Cacciari}, \citenamefont {Salam},\ and\ \citenamefont
  {Soyez}}]{Cacciari:2008gp}%
  \BibitemOpen
  \bibfield  {author} {\bibinfo {author} {\bibfnamefont {M.}~\bibnamefont
  {Cacciari}}, \bibinfo {author} {\bibfnamefont {G.~P.}\ \bibnamefont {Salam}},
  \ and\ \bibinfo {author} {\bibfnamefont {G.}~\bibnamefont {Soyez}},\ }\href
  {\doibase 10.1088/1126-6708/2008/04/063} {\bibfield  {journal} {\bibinfo
  {journal} {JHEP}\ }\textbf {\bibinfo {volume} {04}},\ \bibinfo {pages} {063}
  (\bibinfo {year} {2008})},\ \Eprint {http://arxiv.org/abs/0802.1189}
  {arXiv:0802.1189 [hep-ph]} \BibitemShut {NoStop}%
\bibitem [{\citenamefont {Gaunt}\ \emph {et~al.}(2015)\citenamefont {Gaunt},
  \citenamefont {Stahlhofen}, \citenamefont {Tackmann},\ and\ \citenamefont
  {Walsh}}]{Gaunt:2015pea}%
  \BibitemOpen
  \bibfield  {author} {\bibinfo {author} {\bibfnamefont {J.}~\bibnamefont
  {Gaunt}}, \bibinfo {author} {\bibfnamefont {M.}~\bibnamefont {Stahlhofen}},
  \bibinfo {author} {\bibfnamefont {F.~J.}\ \bibnamefont {Tackmann}}, \ and\
  \bibinfo {author} {\bibfnamefont {J.~R.}\ \bibnamefont {Walsh}},\ }\href
  {\doibase 10.1007/JHEP09(2015)058} {\bibfield  {journal} {\bibinfo  {journal}
  {JHEP}\ }\textbf {\bibinfo {volume} {09}},\ \bibinfo {pages} {058} (\bibinfo
  {year} {2015})},\ \Eprint {http://arxiv.org/abs/1505.04794} {arXiv:1505.04794
  [hep-ph]} \BibitemShut {NoStop}%
\bibitem [{\citenamefont {Boughezal}\ \emph {et~al.}(2015)\citenamefont
  {Boughezal}, \citenamefont {Focke}, \citenamefont {Liu},\ and\ \citenamefont
  {Petriello}}]{Boughezal:2015dva}%
  \BibitemOpen
  \bibfield  {author} {\bibinfo {author} {\bibfnamefont {R.}~\bibnamefont
  {Boughezal}}, \bibinfo {author} {\bibfnamefont {C.}~\bibnamefont {Focke}},
  \bibinfo {author} {\bibfnamefont {X.}~\bibnamefont {Liu}}, \ and\ \bibinfo
  {author} {\bibfnamefont {F.}~\bibnamefont {Petriello}},\ }\href {\doibase
  10.1103/PhysRevLett.115.062002} {\bibfield  {journal} {\bibinfo  {journal}
  {Phys. Rev. Lett.}\ }\textbf {\bibinfo {volume} {115}},\ \bibinfo {pages}
  {062002} (\bibinfo {year} {2015})},\ \Eprint
  {http://arxiv.org/abs/1504.02131} {arXiv:1504.02131 [hep-ph]} \BibitemShut
  {NoStop}%
\end{thebibliography}%

\bibliographystyle{apsrev4-1}
\end{document}